\documentclass[pre,twocolumn,a4paper,showpacs]{revtex4-1}
\usepackage{amssymb}
\usepackage{wasysym}
\usepackage{graphicx}
\usepackage{epsfig}
\usepackage{subfigure}
\begin{document}

\title{The consensus in the two-feature two-state  one-dimensional Axelrod model revisited}

\author{ Elias J.\ P.\  Biral, Paulo F.\ C.\ Tilles and Jos\'{e} F.\  Fontanari }

\affiliation{Instituto de F\'{\i}sica de S\~ao Carlos,
  Universidade de S\~ao Paulo,
  Caixa Postal 369, 13560-970 S\~ao Carlos, S\~ao Paulo, Brazil}
  
  \pacs{89.65.-s, 89.75.Fb, 87.23.Ge, 05.50.+q}

\begin{abstract}
The Axelrod model for the dissemination of culture  exhibits a rich spatial distribution of cultural domains, which depends  on the values of the two model parameters: $F$, the number of cultural features and $q$, the common number of states each feature can assume.  In the 
one-dimensional model with
$F=q=2$,  which is closely related to the constrained voter model, Monte Carlo simulations indicate the existence of multicultural absorbing configurations in which at least one macroscopic domain coexist with a multitude of microscopic ones in the thermodynamic limit. However, rigorous analytical results for the infinite system starting from the configuration where all cultures are equally likely show convergence to only monocultural or consensus configurations.  Here we show that this disagreement is due simply  to the  order that  the time-asymptotic limit   and the thermodynamic  limit   are taken in the simulations. In addition, we show how the consensus-only result  can be derived 
using Monte Carlo simulations of finite chains.
 
\end{abstract}

\maketitle

\section{Introduction} \label{sec:Intro}
The study of  Axelrod's  model for the dissemination of culture  \cite{Axelrod_97} by the statistical physics community has revealed a rich  dynamic behavior   with a nonequilibrium  phase transition separating stationary regimes characterized by distinct distributions of
domain sizes  \cite{Castellano_00,Klemm_03,Vazquez_07,Barbosa_09,Castellano_09}.
Essentially, the different regimes are characterized by the presence or  not of cultural domains of macroscopic size in the thermodynamic limit.
   In Axelrod's   model,   the agents  are represented by  strings  of cultural features of length $F$, where each feature can adopt  $q$  distinct states or traits. Axelrod uses the term culture  to indicate any  set of individual attributes that are susceptible to social influence \cite{Axelrod_97}.

A feature that sets Axelrod's model  apart from most
lattice models that exhibit nonequilibrium phase transitions \cite{Marro_99} is  that  for finite systems all stationary 
states of the dynamics are absorbing configurations, i.e., the dynamics always freezes in one of those configurations. This
contrasts with lattice models that exhibit an active state in addition to a macroscopic number of  absorbing states \cite{Jensen_93}
and the phase transition occurs between the active state and the (equivalent) absorbing states. 
Since according to the rules of Axelrod's model the interaction between two neighboring agents occurs with a probability proportional to
the number of  cultural states they have in common, agents who do not have any cultural state 
in common cannot interact and the interaction between agents who share all their cultural states does not result in any change. Hence  we can guarantee that at an absorbing configuration  any pair of neighbors are either identical  or completely different  regarding their cultural states.  In principle,  Axelrod's model can exhibit monocultural (consensus)  absorbing configurations as well as multicultural absorbing configurations. 

As Axelrod's model can be seen as  $F$ coupled voter models \cite{Ligget_85}, most of the information we have on the behavior of the model in regular lattices was obtained using Monte Carlo simulations of  lattices of finite linear size $L$ and  then properly extrapolating the results  to the thermodynamic limit $L \to \infty$ within a well-established framework in statistical physics (see, e.g., \cite{Privman_90}).
Hence our surprise with  the recent claim by Lanchier \cite{Lanchier_12a} (see also \cite{Lanchier_12b})  that in the  particular case $F=q=2$ of the  one-dimensional system, which is isomorphic to the constrained voter model \cite{Vazquez_03,Vazquez_04}, the Monte Carlo simulations \cite{Vilone_02} yielded predictions that seemed to disagree with his analytical results, leading to the assertions that `spatial simulations are usually difficult to interpret' and that `there is a  need for rigorous  analytical results' \cite{Lanchier_12a}.  In particular, whereas the Monte Carlo results indicate the presence of multicultural absorbing configurations in the thermodynamic limit, Lanchier's analysis shows that only the consensus configurations exist in that limit. Actually, the convergence of the one-dimensional Axelrod's model to a consensus for $q=2$  can be  shown rigorously regardless of the value of $F$ \cite{Lanchier_12b}.

Here  we argue that the reason for
that discrepancy is the  order in which the time-asymptotic limit  $\tau \to \infty$ and the chain size limit $L \to \infty$  are taken in
the simulations. In particular, in the Monte Carlo studies one usually takes the limit $\tau \to \infty$ first and then the limit $L \to \infty$. 
We show that in order to obtain the results of Lanchier's approach, which considers a chain of infinite size at the very outset,  we need to take the limit $L \to \infty$  before the time-asymptotic limit  $\tau \to \infty$ in the Monte Carlo simulations. In doing so, we were able to reproduce numerically  Lanchier's finding that the $F=q=2$ Axelrod model exhibits only a consensus phase in one dimension.

The remainder of the paper is organized as follows. In section \ref{sec:model} we present a brief account of Axelrod's model and point out its connection with the constrained voter model in the case $F=q=2$. The usual order of limits  $\tau \to \infty$ first  and then
$L \to \infty$    is considered in section \ref{sec:equi} and the reverse order in section \ref{sec:finite}. Finally, section \ref{sec:conc} offers our concluding remarks.

\section{Model}\label{sec:model}

In the one-dimensional two-feature two-state    Axelrod's model each agent is characterized by a set of $F=2$ cultural features and each feature   can take on $q=2$  different states, which we label by 0 and 1. Hence there are four distinct cultures
$\left (0,0 \right), \left (0,1 \right), \left (1,0 \right)$ and $\left (1,1 \right)$ in total.
 In the initial configuration each agent is assigned one of these cultures with equal probability. 
The agents are fixed in the sites of a chain of length $L$ with periodic boundary conditions (i.e., a ring).
According to the dynamics of the original model \cite{Axelrod_97},
at each time $\tau$ we pick an agent at random -- the target agent -- as well as one of its neighbors.  As usual in such asynchronous update scheme we choose the time unit as $\Delta \tau = 1/L$. These two 
agents interact with probability equal to  their cultural similarity, defined as the fraction of 
common cultural features. This rule models homophily, which is 
the tendency of individuals to interact preferentially with similar others.  An interaction consists of selecting at random one of the distinct features, and making the
selected feature of the target agent equal  to its neighbor's corresponding state. This rule models social influence since the agents become more similar after they interact.
Hence two neighboring agents with antagonistic cultures
$\left (0,0 \right)$ and $\left (1,1 \right)$ or $\left (0,1 \right)$ and $\left (1,0 \right)$ do not interact, whereas agents with, say,  cultures $\left (0,0 \right)$ and  $\left (0,1 \right)$ can interact with probability $1/2$. In the case the two agents are identical, the interaction produces no changes.  This procedure is repeated until 
the system is frozen into an absorbing configuration.  Clearly, there are four different types of  monocultural absorbing configurations corresponding to each of the four possible cultures, whereas a multicultural absorbing configuration must either be a concatenation of the cultures $\left (0,0 \right)$ and
$\left (1,1 \right)$ or of the cultures $\left (0,1 \right)$ and $\left (1,0 \right)$.

The three-opinion constrained voter model identifies the cultures $\left (0,1 \right)$ and $\left (1,0 \right)$ with a single centrist opinion labeled by $0$ and the other two cultures $\left (0,0 \right)$ and $\left (1,1 \right)$ with a leftist and a rightist opinion
labeled by $-$ and $+$, respectively \cite{Vazquez_03,Vazquez_04}. Leftists and rightists are considered too incompatible to interact so the interactions are between centrists ($0$) and leftists ($-$) or between centrists ($0$) and rightists ($+$) only and follow the usual rules of the voter model \cite{Ligget_85}.   The fact that in the constrained voter model the interaction between, say, $0$ and $-$ takes place with probability 1 whereas in Axelrod's model the interaction between $\left (0,1 \right)$ and $\left (0,0 \right)$ occurs with probability $1/2$ implies only a rescale of time, so that the relaxation  in Axelrod's model takes twice as long as in the constrained voter model.  In addition, the centrist consensus (i.e., the extinction of both  leftists and rightists)  should be interpreted either as the consensus of one of the cultures $\left (0,1 \right)$ and $\left (1,0 \right)$ or as the multicultural coexistence of those two cultures.   As expected, Monte Carlo simulations of
the constrained voter model yielded multicultural coexistence between the two extremist opinions as well as  consensus of one of the three opinions \cite{Vazquez_03}, as in Axelrod's model \cite{Vilone_02}.

\section{Monte Carlo study of the absorbing configurations}\label{sec:equi}

As pointed out already, the main appeal of Axelrod's model to the statistical physics community  is probably  the existence of a phase transition 
 that separates absorbing configurations, which differ in the statistical organization of their cultural  domains,  in the space of parameters $\left ( F,q \right )$ 
 \cite{Castellano_00,Vilone_02,JEDC_05}.  Hence the typical statistical mechanics analysis of Axelrod's model consists of taking first the limit $\tau \to \infty$ for finite $L$, so one is guaranteed to reach the absorbing configurations,  and then
extrapolating the results of finite $L$  to the thermodynamic limit. For the sake of completeness,  in this section we  present the results of the analysis of the absorbing configurations of Axelrod's model for $F=q=2$, expanding upon the study of Vilone et al. \cite{Vilone_02}.
 
\begin{figure}[!h]
\centering\includegraphics[width=0.48\textwidth]{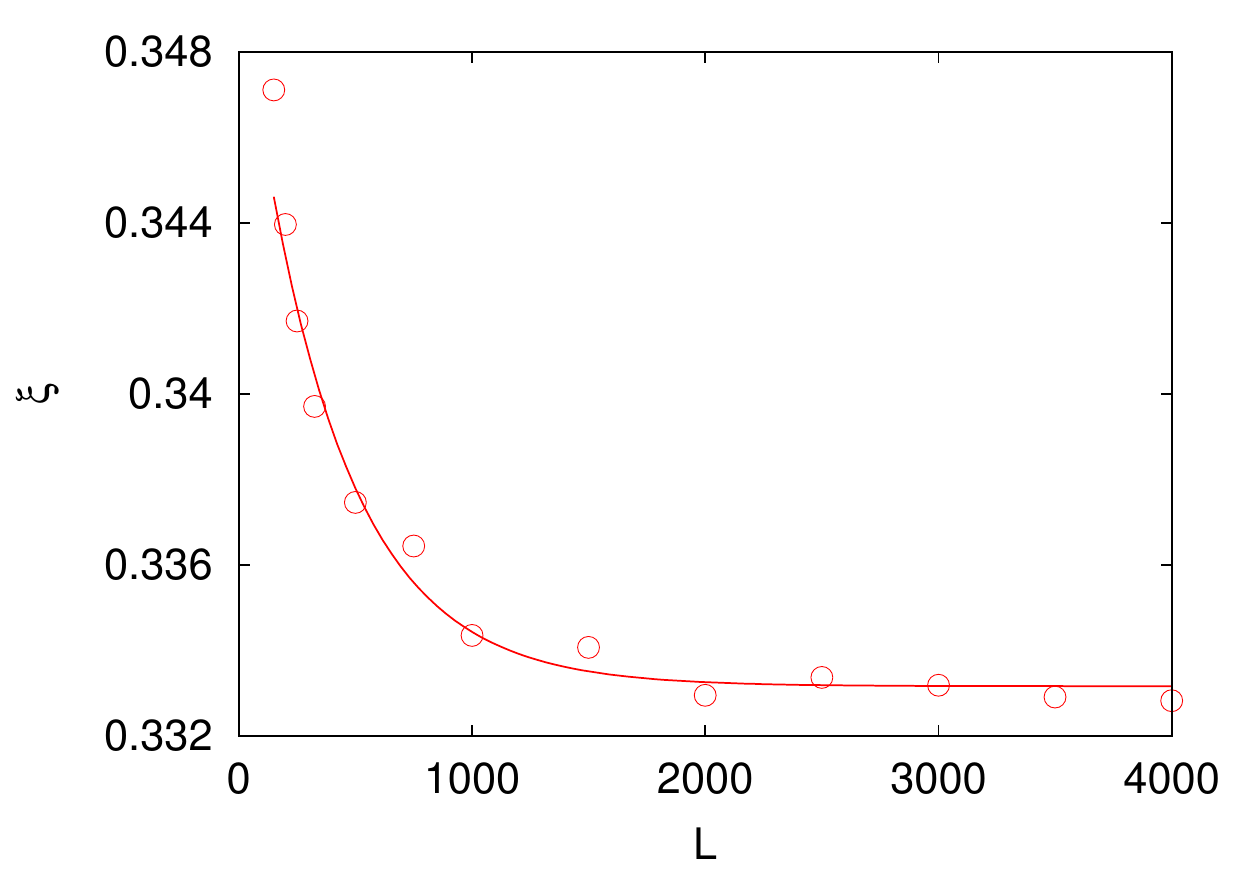}
\caption{Fraction of runs  trapped in monocultural absorbing configurations $\xi$  as function of the chain size $L$. The total 
number of runs is $10^6$  for each $L$. 
The solid line is the fitting $ \xi = 0.33 + 0.017 \exp \left ( -L/387 \right ) $. The asymptotic value of $\xi$ is very robust  whereas the other two adjustable parameters can vary considerably with changes in the range of the fitting.}
\label{fig:1}
\end{figure}
 
A remarkable result about the  case $F=q=2$ is that  when the four cultures are present in the same proportion in the initial configuration, a  fraction $\xi$ of runs freezes in monocultural absorbing configurations whereas the remaining fraction $1- \xi$ freezes in
multicultural configurations.  The dependence of $\xi$ on $L$, which is shown in Fig.\ \ref{fig:1},   is very weak and we found that $\xi \to 0.33$ exponentially fast with increasing $L$. 

\begin{figure}[!h]
\subfigure{\includegraphics[width=0.48\textwidth]{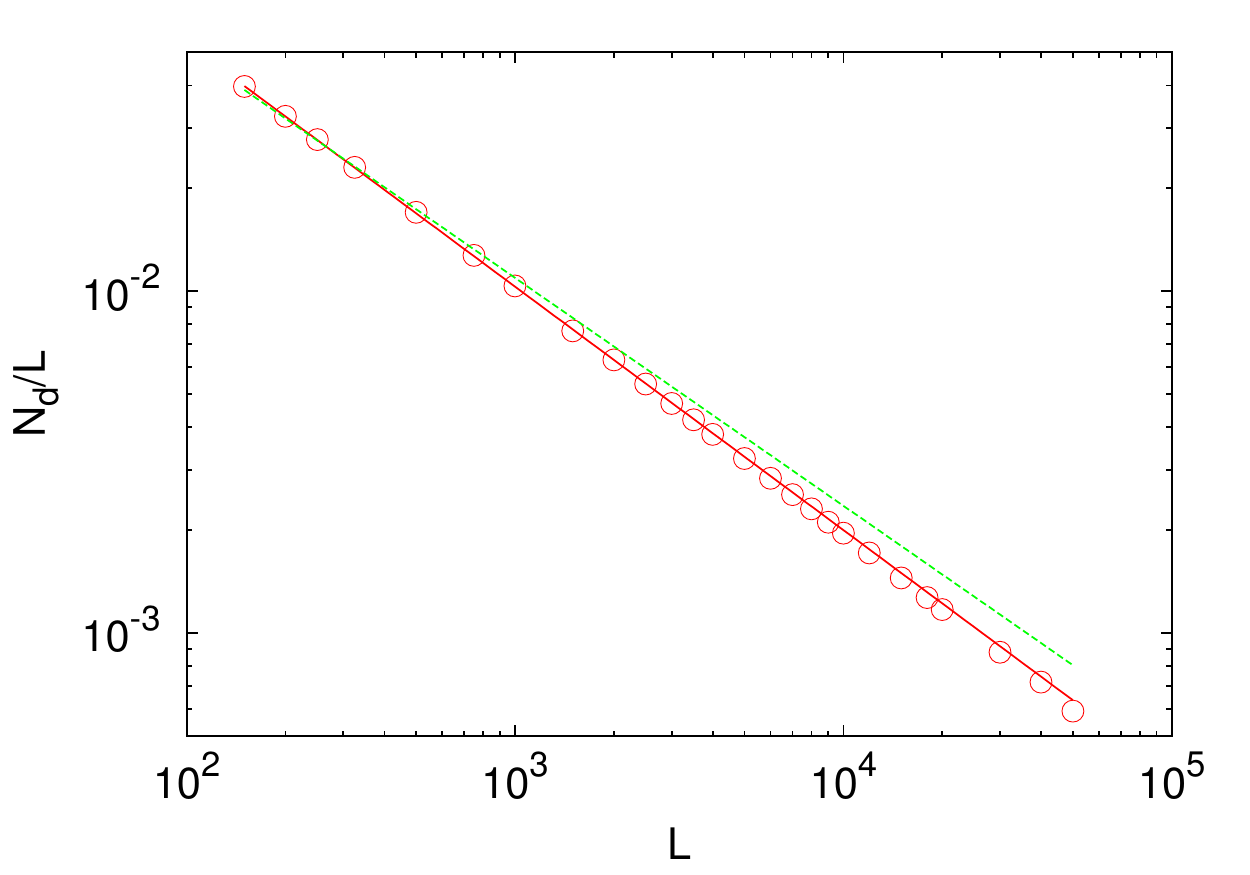}}
\subfigure{\includegraphics[width=0.48\textwidth]{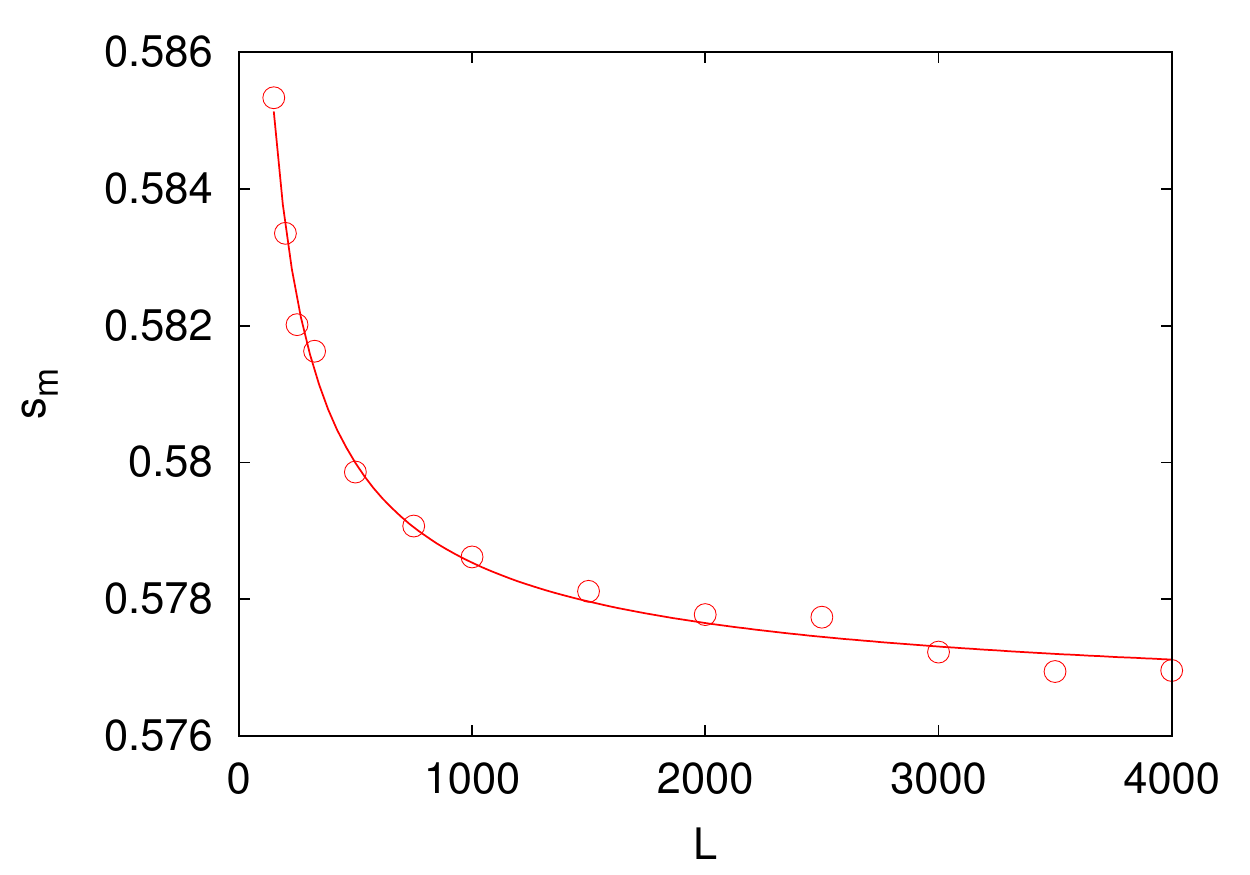}} 
\caption{Upper panel: Mean density of domains $N_d/L$   as function of the chain size  $L$ for the multicultural absorbing configurations. 
The solid line is the fitting $ N_d/L = 1.41 L^{-2 \psi} $ with $\psi = 0.36$ and the dashed line
is the fitting  when the exponent is held fixed at $\psi = 1/3$ (see section \ref{sec:conc}). Bottom panel: Mean fraction of sites $s_{m}$ that belong to the largest
cultural domain as function of the chain size  $L$ for the multicultural absorbing configurations. The solid line is the fitting
$s_m = 0.576 + 0.33/L^{0.72}$. The asymptotic value of $s_m$ is very robust to changes in the fitting function or fitting range.}
\label{fig:2}
\end{figure}

In order to understand the nature of the multicultural absorbing configurations we present in Fig.\ \ref{fig:2} the mean density of domains $N_d/L$  and the mean fraction of sites that belong to the largest domain $s_m$. 
These results  show that $N_d$ increases with $L$ according to the  power law  $L^{1-2\psi}$ with $\psi = 0.36$  and that $s_m \to 0.576$ in the thermodynamic limit, which indicates a rich multicultural equilibrium regime characterized by
the coexistence of at least one macroscopic domain with a large number of  microscopic ones.  
This conclusion is in stark contrast to the claim made by Lanchier  \cite{Lanchier_12a} that Axelrod's model reaches consensus (i.e., converges to a monocultural equilibrium) for $F=q=2$ in the thermodynamic limit.
We stress that only runs that have frozen in multicultural configurations were considered in the calculation of the averages exhibited  in 
Fig.\ \ref{fig:2}. This is the reason why our estimate for $s_m$  differs   from that of Vilone et al. \cite{Vilone_02}, which represent averages over all runs. Of course, their results are  easily derived from ours. For instance, averaging over all runs yields
$\hat{s}_m = \xi + \left ( 1 - \xi \right) s_m \to 0.717$ for $L \to \infty$.

\begin{figure}[!h]
\centering\includegraphics[width=0.48\textwidth]{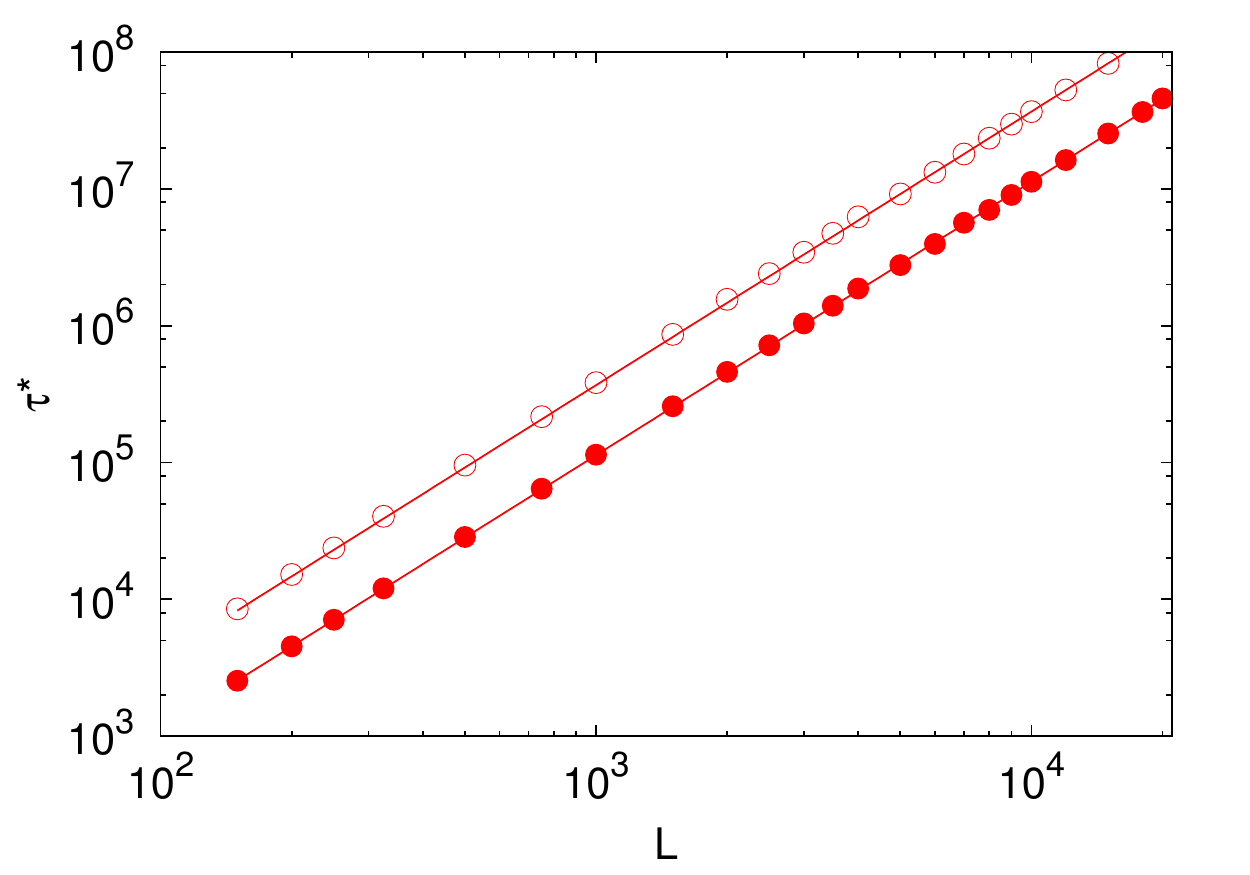}
\caption{Mean relaxation time $\tau^*$ as function of the  chain  size $L$ for runs frozen in monocultural configurations (empty symbols) and for runs frozen in multicultural configurations (filled symbols).  The solid lines are the fittings $\tau^* = 0.37 L^2$ (monocultural)
and $\tau^* = 0.11 L^2$ (multicultural).}
\label{fig:3}
\end{figure}

The reason for the stern discrepancy between the results of the  Monte Carlo simulation and  Lanchier's analysis becomes apparent when we study how the  relaxation time $\tau^*$ scales with the chain size $L$. Figure \ref{fig:3} shows that regardless of the nature of the absorbing configuration we find  $\tau^* \sim L^2$. 
In addition, these results show that the dynamics takes about 3.4 times as long to freeze in a monocultural absorbing configuration than in a multicultural one.

\section{Monte Carlo study of  the finite time dynamics}\label{sec:finite}

The previous section summarized the Monte Carlo findings for  the stationary regime of Axelrod's model with $F=q=2$, 
where  the limit $\tau \to \infty$ is taken keeping  the chain size $L$ fixed. To obtain the results of Lanchier \cite{Lanchier_12a} we have  to take the limit $L \to \infty$ for a  fixed time $\tau$. Here we focus on two critical measures of the dynamics. The first measure is  the density of bonds (links or edges) that connect two sites that have no features in common. Since those sites do not interact we refer to the bonds connecting them as walls and denote their density by $n_F$. The second measure is the density of bonds that connect sites with exactly one feature in common. Since those sites can interact we refer to those bonds as active bonds and denote their density by $n_A$. In doing so we  conform to the terminology of Vilone et al. \cite{Vilone_02}, in which $n_F$ stands for  the density of
bonds connecting sites that differ by all $F$ features.  In the terminology of Lanchier \cite{Lanchier_12a}, the measures $n_F$ and $n_A$  are the densities of 0-edges and 1-edges, respectively, which were shown to vanish in the
asymptotic limit $\tau \to \infty$ for a chain of infinite size. We note that for the one-dimensional lattice with periodic boundary conditions (i.e., a ring)  the density of walls $n_F$ becomes identical to the density of domains $N_d/L$ when the dynamics freezes in
the absorbing configurations.

\begin{figure}[!h]
\centering\includegraphics[width=0.48\textwidth]{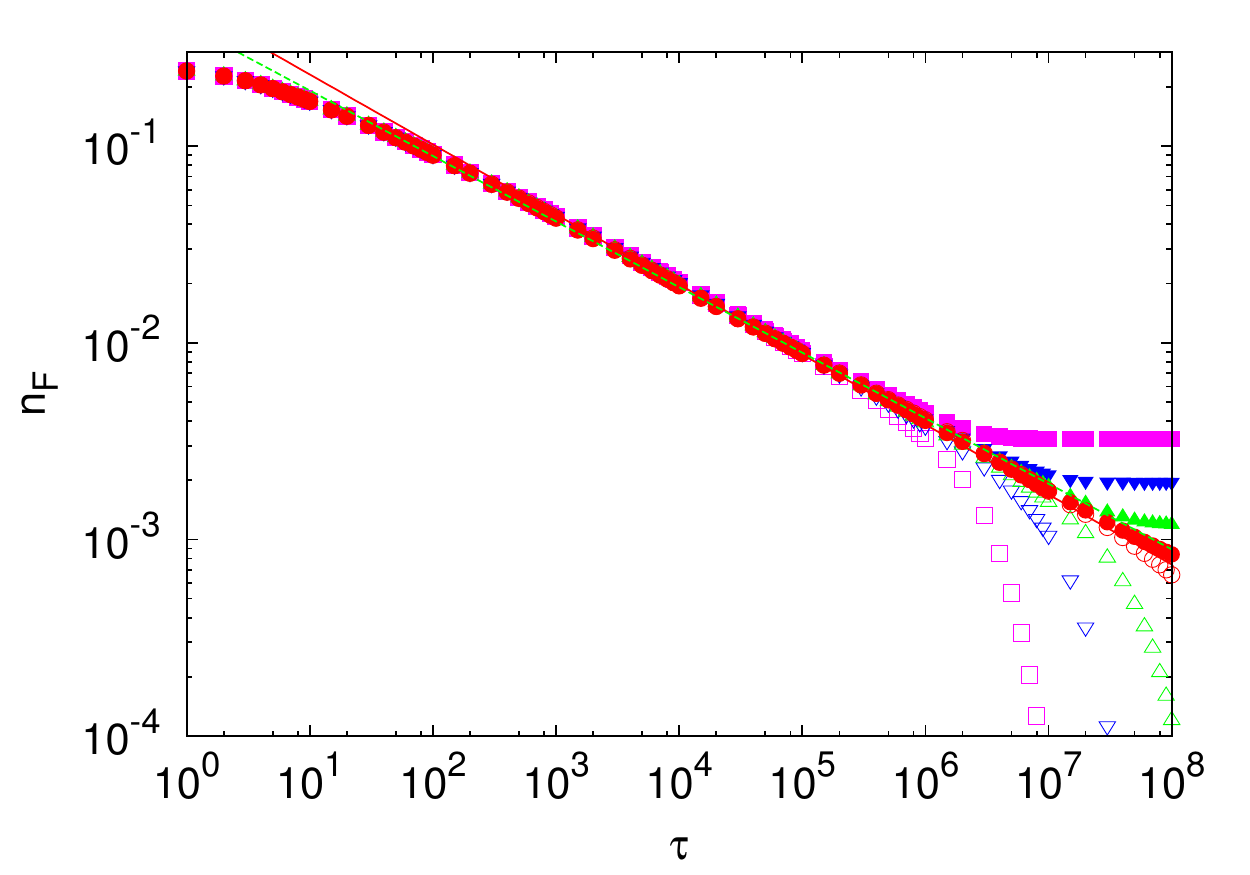}
\caption{Density of walls $n_F$  as function of  the time (Monte Carlo steps) $\tau$      for chains of size
$L = 5 \times 10^3 (\Box), 1 \times 10^4 (\triangledown), 2 \times 10^4 (\triangle)$ and $5 \times 10^4 (\circ)$. The open symbols represent 
averages taken over runs that converged to monocultural (consensus) absorbing configurations, whereas the filled symbols represent the statistics over the runs that converged to multicultural configurations. 
The solid line is the fitting $ n_F = 0.52 \tau^{-\psi} $, with $\psi = 0.36$, of the data for $L=5 \times 10^4$  in the range $\tau \in \left [ 10^3,10^7 \right ]$. The dashed line is the fitting  with the exponent held fixed at $\psi = 1/3$ (see section \ref{sec:conc}) }
\label{fig:4}
\end{figure}

\begin{figure}[!h]
\centering\includegraphics[width=0.48\textwidth]{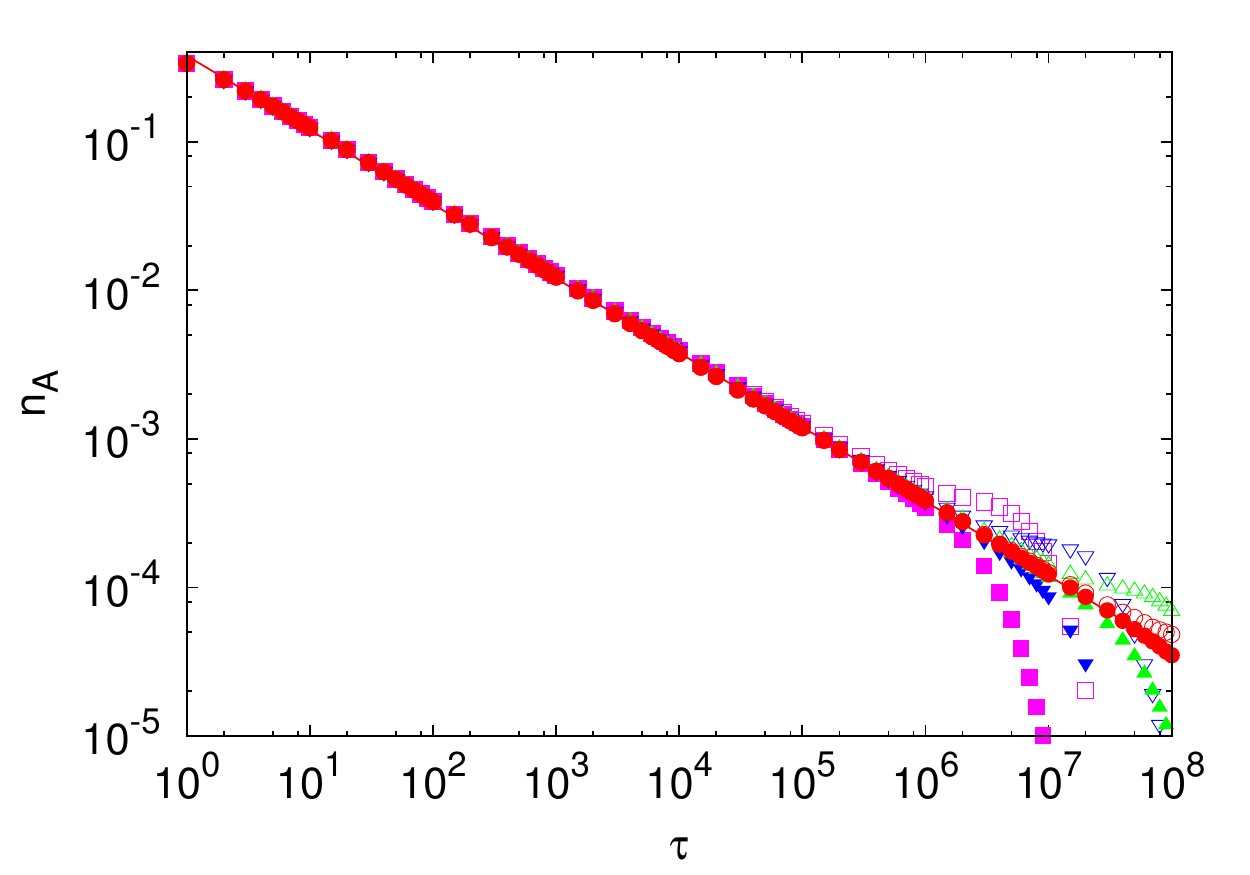}
\caption{Density of active bonds $n_A$  as function of the  time  $\tau$. 
The symbol convention is the same as in  Fig.\ \ref{fig:4}.  
The solid straight  line is the fitting $ n_A = 0.38 \tau^{-1/2} $  in the region where   $n_A$ 
is insensitive to the chain size.}
\label{fig:5}
\end{figure}

Figures \ref{fig:4} and \ref{fig:5} show the time evolution of the density of walls and active bonds, respectively, for several  values of chain size $L$. For each $L$ there are two sets of data, according to whether the dynamics freezes in a monocultural absorbing configuration (empty symbols) or to  whether  it freezes in a multicultural configuration (filled symbols).  In agreement with the results of the previous section, for any finite $L$ the statistics over the runs leading to multicultural absorbing configurations results in a nonzero value  for $n_F$, even in the limit  $\tau \to \infty$. Of course, for the runs leading to monocultural absorbing configurations we find that $n_F \to 0$ even for finite $L$. The point here is that unless $\tau$ is on the order of $L^2$ it is not possible to distinguish between those two cases. More importantly,  we can immediately realize what happens in the case that $L \to \infty$ with $\tau$ finite:  when the limit $L \to \infty$
is taken before the limit $\tau \to \infty$, all data fall on the  power law function $n_F \sim \tau^{-\psi}$, with $\psi = 0.36$ for large $\tau$ (solid line in Fig.\ \ref{fig:4}),
from where we conclude that $n_F \to 0$ as $\tau \to \infty$ in agreement with Lanchier \cite{Lanchier_12a}.

Regarding the density of active bonds $n_A$ shown in Fig.\ \ref{fig:5}, this quantity tends to zero for finite $L$ and large $\tau$ irrespective of the nature of the absorbing configuration. In particular, in the limit $L \to \infty$ we find $n_A \sim \tau^{-1/2}$ for large $\tau$. This figure reveals also that the relaxation towards monocultural absorbing configurations is much slower than towards multicultural configurations, in agreement with the results of Fig.\ \ref{fig:3}. However, the additional information Fig.\ \ref{fig:5} offers is
that the slowing down takes place near the end of the runs when the density of active bonds  levels off before resuming its decrease
towards zero. 

It is important to note that Vazquez et al. \cite{Vazquez_03}  have calculated the relaxation of $n_A$ exactly by mapping the constrained voter model on a spin-$1/2$ ferromagnetic Ising chain with zero-temperature Glauber dynamics \cite{Glauber_63}. Most interestingly,  those authors used the asymptotic decay of $n_A \sim \tau^{-1/2}$, which is due to the 
underlying diffusive dynamics,  to obtain the scaling of the mean relaxation time $\tau^* \sim L^2$ since this is the typical time needed for the active bonds to diffuse throughout the chain and be eliminated \cite{Vazquez_03}.  However, since that mapping, applies to the dynamics of $n_A$ only the relaxation of $n_F$  has to be studied through Monte Carlo simulations.

To conclude this section, we note that $ \mathcal{L} = 2n_F + n_A$ is a Lyapunov function  for  the one-dimensional Axelrod model with $F=q=2$ and periodic boundary conditions  \cite{JEDC_05}. Since the dynamics never increases the value of $\mathcal{L}$,
the consensus configurations are  global minima of this function ($\mathcal{L}=0$), whereas the multicultural configurations are local minima ($\mathcal{L}=2n_F > 0$) for finite systems. As a result, the multicultural absorbing configurations are unstable to small local perturbations
(cultural drift), which then drive the system towards one of the consensus configurations.

\section{Conclusion}\label{sec:conc}

We show that the reason for the discordance between the Monte Carlo simulations of the two-feature two-state one-dimensional
Axelrod model \cite{Vilone_02} (or, equivalently, the  constrained voter model \cite{Vazquez_03})  and the rigorous analytical results of Lanchier for infinite chain sizes \cite{Lanchier_12a} is that the limits $\tau \to \infty$ and $L \to \infty$ do not commute in the Monte Carlo simulations. This difficulty should be  expected somehow: since the mean relaxation time  scales with $L^2$, taking the limit $L \to \infty$ with finite $\tau$,  as we have done here in order to obtain  Lanchier's results,  will keep the system away from the stationary regime no matter how large  one chooses the value of $\tau$. 

Actually, the fact that the order in which the limits $L \to \infty$ and $\tau \to \infty$ are taken influences the results was already explicit in the scaling form proposed by Vilone et al. for the mean density of walls,
\begin{equation}\label{scal}
n_F \left ( \tau, L \right)  = \tau^{-\psi} g_F \left ( \tau/ L^2 \right )
\end{equation}
where the scaling function is such that  $g_F \left ( x \right ) = const$ for $ x \ll 1$ and $g_F \left ( x \right )\sim x^{\psi} $ for $x \gg 1$ \cite{Vilone_02}. Here $n_F$ represents an average over all runs or, equivalently for our purposes, over  the runs  trapped in multicultural absorbing configurations.  For those runs we recall from Fig.\ \ref{fig:2}  that $n_F \sim N_d/L \sim L ^{-2\psi}$ for $\tau \gg L^2$.  Vilone et al. estimated $\psi=1/3$ using chains of size up to $L=5 \times 10^3$.  While this estimate is practically indistinguishable from ours (i.e., $\psi = 0.36$) in the scale of Fig.\ \ref{fig:4}, it is clearly inaccurate when we consider the equilibrium regime 
$n_F \sim N_d/L$ shown in Fig.\ \ref{fig:2}. We point out, however,  that the value of the exponent $\psi$ does not seem to reflect any meaningful property of the  model since it varies with the initial frequency of cultures, as shown in the context of the constrained voter model \cite{Vazquez_03}.

A word is in order about the behavior of the one-dimensional Axelrod model for  more general values of the parameters $F$ and $q$. For
$F=2$ and $q>2$  the dynamics of the infinite system converges to highly fragmented multicultural configurations \cite{Lanchier_12a,Vilone_02} and the mean-field expression $n_F \left ( \tau \to \infty \right) = 1 - 2/q$ fits perfectly the numerical results
\cite{Vilone_02}. For $q=2$ and $F>2$ the dynamics of the infinite system converges to a consensus \cite{Lanchier_12b}, in agreement with
the simulations \cite{Vilone_02}. For  $F> 2$ and $q > 2$, the Monte Carlo simulations indicate that  the infinite system exhibits a discontinuous transition between
a consensus phase that exists for small $q$ and a multicultural phase that exists for large $q$ \cite{Vilone_02}.
The  discordance between the full mathematical analysis of  the infinite system and the Monte Carlo simulations discussed at length in this paper occurs for $F=q=2$ only, and provides a simple example of a situation where a rigorous solution of a model in the thermodynamic limit can miss important facts that necessarily would appear in applications to finite systems \cite{Toral_07}.

In sum, the scaling form (\ref{scal}),  which  is validated by  the Monte Carlo simulations,  explains how the order of the
limits $\tau \to \infty$ and $L \to \infty$ determines the nature of the absorbing configurations in the two-feature two-state one-dimensional Axelrod model and shows that  there is really no contradiction between the Monte Carlo  and Lanchier's results.

\acknowledgments
The work of J.F.F. was partially supported  
 by  grant    2013/17131-0, S\~ao Paulo Research Foundation (FAPESP)
and by grant  303979/2013-5, Conselho Nacional de Desenvolvimento Cient\'{\i}fico e Tecnol\'ogico (CNPq).
P.F.C.T. was supported by grant  2011/11386-1, S\~ao Paulo Research Foundation (FAPESP). E.J.P.B. is
supported by a CAPES fellowship.

\end{document}